\documentclass[a4paper,12pt]{article}
\usepackage[cp1251]{inputenc}
\usepackage{graphics}
\usepackage{amssymb,amsmath}
\usepackage[english]{babel}
\usepackage{indentfirst}
\begin{document}

\bigskip
\centerline {\bf Spotting in stars with a low level of activity,
cloze to solar activity}

\bigskip

\centerline {E.A.Bruevich $^{a}$, I.Yu.Alekseev$^{b}$}

\bigskip

\centerline {Sternberg Astronomical Institute, Moscow, Russia}\
\centerline  {Scientific Research Institute Crimean Astrophysical
Observatory, Ukraine}\

\centerline
{E-mail:$^a${red-field@yandex.ru},$^b${ilya@crao.crimea.ru}}\

\bigskip

  Data on the variability of the continuum optical emission are used for the first time to estimate
  the degree of spotting in stars similar to that of the sun. It is sown that the amount of spotting
  increases gradually from the sun to the highly spotted stars for which Alekseev and Gershberg constructed
  the zonal model for the distributions of spots. A close relationship is found between spotting and the
  power of the x-ray emission from the stars with widely varying levels of activity


\vskip12pt {\it\bf 1. Introduction}

The optical photometric observations of red dwarf stars regularly
made at various observations since the beginning of the 1950's have
revealed a low-amplitude quasiperiodic variability in the brightness
of some of these stars owing to the presence of dark spots similar
to sunspots on the surface of rotating stars. In turns out that the
observed brightness variations are mainly related to the fact that
the total spot area of the most highly spotted stars can be
substantially greater than on the sun at the cycle maximum,
approaching up to 40 \% of the disc surface. Further observations
have confirmed the existence of all the basic manifestations of
solar-type activity on these stars, at the photospheric as well as
at the chromospheric and coronal levels [1]. The chromospheric
activity of stars has been studied for more than 40 years in a
special program of regular observations of the H and K lines of Ca
II, initially at Mount Wilson Observatory and then at the
Smithsonian Observatory. Several of 111 stars from this so-called
"HK-project" have regular cyclical variations in their chromospheric
radiation with periods ranging mostly from 7 to 15 years [2]. The
stars have been divided into groups based on the characteristics of
their long-term brightness variations and how distinctly the cycles
are expressed [2].

The launching of the High Astrophysical Observatory HEAO-2 in 1978
triggered extensive studies (in soft x-ray range) of the coronal
activity of the stars of late-type spectral class.

The ROSAT satellite [3] has contributed the most data on the x-ray
emission of these late-type stars. It turns out the best indicator
of activity in the outer atmospheres of stars in the ratio of the
x-ray and bolometric luminosities, $L_x/L_{bol}$. This index is
currently known for several hundred stars. The highest values of
$L_x/L_{bol}$ are found in stars of spectral class F with a high but
irregular activity [4].

Recently several stars have been found to have cyclical variations
in their soft x-ray fluxes, which confirms the existence of cyclical
activity of the solar type stars of F,G, K and M spectral classes
(photospheric, as well as chromospheric and coronal activity).

In this paper we limit ourselves only to examining quasistationary
effects - the development of active regions, while neglecting
flares, as such.

There are several ways or determining the relative area of spots
from observations of long-term variability in the continuum or in
the lines in the optical range. The most widespread methods are:
constructing charts of surface nonuniformities (Doppler mapping or
imaging) and the zonal model method. In most cases these two methods
yield similar values for the spot areas on the surface of stars.
Zonal model have been constructed for the stars whose photometric
variability have been regularly observed at the CrAO (Crimean
Astrophysical Observatory), and the area of the dark spots on their
surfaces has been determined quite accurately [5,6]. This sample of
stars includes several well known M-dwarfs (flare stars), as well as
about a dozen K and G-stars. The amplitude of the rotational
modulation in the optical continuum of these stars shows up
distinctly, so that the amounts of spotting on their surfaces are
substantial. We shall refer to this sample of stars as highly
spotted.

In the course of a comparatively recent program of searching for
planets, about one thousand stars with solar-type activity have been
discovered [7]. These one thousand stars with observed emission in
the H and K Ca II lines include a rather large number of stars for
which the level of chromospheric activity is fairly low, similar to
that of the sun. The spotting of the surface of these stars had to
be determined in order to compare it with other characteristics of
their activity. Photometric observations of some of these stars with
a weak surface activity have been made over the last 20 years as a
part of the "HK-project"{}[8].

In this paper we study the spottedness of late type stars that have
different levels of activity and compare this data to other
characteristics of their activity.

\vskip12pt {\it\bf 2.Estimating the spottedness of solar-type stars}

The "HK-project" observational programm for studding the
chromospheric activity of stars includes determination of the ratio
of the radiative flux at the centers of the H and K emission lines
of Ca II ($396,8$ and $393,4$ nm respectively) to the continuum flux
($400,1$ and $390,1$ nm), that is so-called S Ca II index [2].

S Ca II index or S-index is average value over the two Ca II lines
($396,8$ and $393,4$ nm) normalized on the continuum flux (average
value over fluxes in $400,1$ and $390,1$ nm).

In the following we compare the spottedness of two groups of stars:
highly spotted, which have been studied at the CrAO, and lowly
spotted stars from the "HK-project"{}, which are closed to sun.

To do this we have used photometric observations data in the
Stromgren "b" and "y" bands [8]; the (b + y)/2 values are very close
to the standard Johnson V band. The group of "HK-project"{}stars
with clearly distinct emission in the Ca II lines belonging to
spectral classes F, G and K, which rarely overlap the sample of
highly spotted stars, included only a few late K and one M star. In
our investigation there are only two stars common to both groups.

In order to estimate the spottedness of the solar-type stars, we
used the above mentioned photometric observations of 35 stars in the
"HK-project"{}[8]. These 35 stars were observed over 20 years in
parallel with an extension of the almost 40-year series of
observations of the same stars in the H and K lines among 111
others. The modulation in the photometric light curves of the
"HK-project"{} stars was, on the average, much lower than for the
stars observed at the CrAO. This is understandable, since the stars
observed at the CrAO and the "HK-project"{} stars belong to
different samples of stars: the former are the brightest of spotted
stars while the latter are characterized by exceptional
chromospheric activity.

All the more interesting is a comparison of these stars from the
standpoint of their simultaneous manifestation of spotting and of
chromospheric and coronal activity. In our case, our observations in
a single photometric band lead to an estimate for the overall
spottedness of the stellar surface from the amplitude of the
rotational modulation and the maximum brightness of the star
corresponds to the level of the unspotted photosphere. We use an
expression for the brightness of a spotted star based on the Vogt
approximation [6],

$$ \Delta m_{\lambda}
=-2.5\log(1-(1-\beta_{\lambda})G_{\lambda}),     \eqno(1)   $$
where
$$ G_{\lambda}=((1-u_{\lambda})I+u_{\lambda}J)/(I-u_{\lambda}/3)    \eqno(2)  $$

Here the contrast $ \beta_{\lambda} $ - is the ratio of the surface
brightness of a spot and the photosphere,

  $ u_{\lambda} $ -  is the linear limb darkening coefficient,
$I$ is the area of the projection of the spot on the plane on the
chart expressed as a fraction of the star's visible disk, and $J$
characterizes the position of the spot relative to the center of the
disk. The difference in stellar magnitudes ,
 $ \Delta m $, is taken relative to the brightness level of the unspotted
 atmosphere. If the limb darkening is neglected, then approximately
$ J=2I/3$, and we obtain the following relation for calculating the
fraction of the disc occupied by spots:

$$ I =(1- 10^{-0.4\Delta m})/(1-\beta_{\lambda})   \eqno(3)  $$

Using Eq. (3) for for the "HK-project"{} stars, we obtain typical
estimates for $I$ which are numerically equal to the relative area
of the stars disk occupied by spots $S_{spot}$. We determine the
corresponding values of
 $ \Delta m $ from the photometric light curves for these stars in
 Ref. 8.
One of the basic results of the zonal model analysis of the
observations was the discovery of a distinct coupling between the
temperature of the spots and absolute brightness $M_V$ of a star
(Fig. 1).

We used this relationship to determine the contrast
 $\beta_{\lambda} $
 for the other late spectral class stars, as well, in particular for
 the "HK-project"{} stars discussed here. The absolute brightness of
 these stars is known and have taken it from Ref. 8.

 The Figure 1 shows that the dependence of the relative spots area $S_{spot}$
on the contrast values $\beta_{\lambda} $ varies negligible over
 $ \beta_\lambda = 0.1 - 0.25$. The feasibility of estimating the
 spottedness of the "HK-project"{} stars using a simplified
 involving a determination of the contrasts
$ \beta_{\lambda} $ of the stars as a function of the absolute
stellar magnitude $M_V$ has been checked by means of revers
calculations of the overall area of the spots for 25 of the CrAO
stars (whose spottedness is known from zonal model calculations)
using Eq. (3). The greatest error in our calculations occurred in
the case of the star EK Dra = HD129333, which is common to both
groups of stars, where the amount of spottedness according to zonal
model is a factor 1.2 - 1.3 times exceeds the value we determined.
The spottedness is the same for the both methods in the case of the
second star that is common to both groups of stars, BE Cet = HD1835.

We also analyzed the dependence of the area of maximum spottedness
for 24 stars from "HK-project"{} (for these stars there appear data
from ROSAT x-ray catalog) and for the 24 stars observed at the CrAO
at the color index $(B-V)$. It turns out that stars with active
chromospheres, with greater spottedness than the other
 objects in the "HK-project"{}, belong mainly to spectral classes
 F,G, and, partly, K. At the other hand most of the highly spotted
 CrAO stars belong to spectral classes K and M. Clearly, the "HK-project"{}
 stars whose activity level is close to that of the sun have
 considerably less spotting. In addition, the portion of these stars
 with well determined cycles of chromospheric activity, including
 the sun, are characterized by a still lower (by a factor of 2-3)
 spot area (the only exception is V2292 Oph=HD152391) compared to
 the other stars in the "HK-project"{} that were considered, whose
 spottedness is, in turn, a factor 2-5 times less that of stars
 observed at the CrAO.

Data on soft x-ray emission from the ROSAT satellite exists for 46
stars in two groups that were examined (EK Dra and BE Cet are common
to the both groups). Along with determining the degree of spotting
of the stars, $S_{spot}$ (the relative area of the visible disk of
the star occupied by and expressed as a percent), we analyzed the
relationship between $S_{spot}$ and $L_x/L_{bol}$. As note above,
 $L_x/L_{bol}$ is a good index for characterizing the power of the
 stellar coronae.

Our calculations of the maximum spotted area for the 24
"HK-project"{} stars chosen for fuhrer analysis and for the 24 stars
observed at the CrAO are listed in Table.

The second column in Table consists of the bolometric luminosities $
L_{bol}$. These values of the "HK-project"{} stars are taken from
Ref. 4 and the method employed there we used to calculate $L_{bol}$
for the CrAO stars.

The fifth column in Table - the values of spottedness $S_{spot}$ of
stars were calculated using Eq. (3) for the "HK-project"{} stars
(rows from 1 to 23). The values of spottedness $S_{spot}$ of stars,
observed in CrAO (rows from 24 to 46) were taken from zonal model
calculations [5,9].

The values of spottedness $S_{spot}$ of stars
 as plotted as a function of the
ratio $\log(L_x/L_{bol})$ are shown in Fig. 2.

Here it was found for the first time that the amount of spotting
varies gradually from the sun to the stars with the most powerful
coronae (of the individual stars). Clearly, the gradients in the
plot $S_{spot}$ as a function of $\log(L_x/L_{bol})$ for solar-type
stars and highly spotted stars are quite different. This is because
of some of the highly spotted stars are already at the saturation
level for x-ray emission, when a fraction 0.001 of the energy
generated in the core of the star is expended in heating the
coronae. Of course, the portion of the objects which are quieter
than the sun were not included in our examination and should lie in
the lower left hand corner of Fig. 2.

It should be noted that the method proposed here for determining the
contrasts of the spots from zonal model calculations (Fig. 1),
rather than from observations, does introduce some error into the
determination of the amount of spottedness, but this error is small,
because the contrasts vary insignificantly from star to star and
only influence the result indirectly. There may also be some error
in determining the brightness of stars "without spots" because we
have used photometric observations of "HK-project"{} stars spanning
an time interval of 20 years. Thus, there is some probability that
we have neglected long-period variations in the brightness over a
longer or equal to 20 year time scale. Some of this discrepancy in
the spottedness determination for EK Dra may be related to just
factor.

\begin{figure}[h!]
 \centerline{\includegraphics{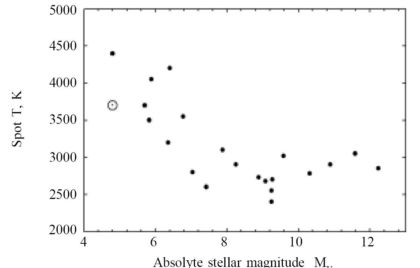}}
 \caption{The temperature of the spots as a function of the absolute stellar magnitude according to zonal model
 calculations [6]. The symbol $\odot$ denotes the sun}\label{Fi:Fig1}
\end{figure}

\vskip15pt {\it\bf 3. Conclusions}

More than one thousand stars with processes similar to solar
activity are currently known in the neighborhood of the sun. This
makes it possible to investigate how the characteristics of the
activity vary on going from stars with a low level of activity
similar to that of the sun, to stars with soft x-ray emission with
level of atmospheric activity close to maximum level of all the
possible (close to saturation).

In this paper we have determined the amount of spotting of the
surface of low-activity stars from the long-term variability in the
optical continuum. It has been found that the calculated spotting
area increases from the solar value of 0.2 \% at the 11-year cycle
maximum, to 1-5\% for stars in the "HK-project"{}, and than
increases sharply to 20-35 \% in highly spotted stars, whose
spotting areas have been determined from observations and zonal
model calculations at the CrAO.

\begin{figure}[h!]
 \centerline{\includegraphics{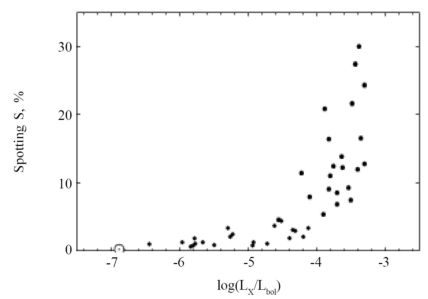}}
 \caption{The area of the spotted regions as a function of the ratio $\log(L_x/L_{bol})$The symbol $\odot$ denotes the sun}\label{Fi:Fig2}
\end{figure}

A comparison of the spottedness with the relative fraction of energy
expended in heating the coronae of the stars  revealed a rather
close coupling of  $ S_{spot}$ and $lg(L_x/L_{bol})$ values. It has
been suggested earlier that the deficit of radiation which is not
radiated in the spots in the optical range is expended in changing
the outer atmosphere of active stars of F, G, K and M spectral
classes [9]. The analysis given in this paper shows that this idea
is realized to the greatest extent in the coronae. Specifically,
spot formation leads primarily to the formation of high temperature
regions in the corona and, accordingly, to a substantial rase in the
x-ray fluxes. The sun fits well into this pattern, although its
level of spottedness and x-ray emission are substantially lower than
in other stars with stable activity cycles.

\begin{center}
 \begin{tabular}{|c|c|c|c|c|}
   \hline
  HD Number, Star's Name         &$\log(L_x)$ & $\log(L_{bol})$ & $\log(L_x/L_{bol})$ & $S_{spot}$, \% \\ \hline
  SUN                            & 26,7       & 33,58           & -6,88               & 0,20           \\ \hline
  HD81809                        & 28,1       & 33,59           & -5,49               & 0,80           \\ \hline
  HD114710,$\beta$ $Com$         & 28,06      & 33,72           & -5,66               & 1,20           \\ \hline
  HD115404                       & 28,02      & 33,28           & -5,26               & 2,00           \\ \hline
  HD160346                       & 27,48      & 33,25           & -5,77               & 1,00           \\ \hline
  HD201091, $61 CygA$            & 27,15      & 33,1            & -5,95               & 1,20           \\ \hline
  HD201092, $61 CygB$            & 27,15      & 32,93           & -5,78               & 1,80           \\ \hline
  HD149661,$V2133 Oph$           & 28,16      & 33,38           & -5,22               & 2,40           \\ \hline
  HD1835,$BE Cet$                & 28,99      & 33,54           & -4,55               & 4,50           \\ \hline
  HD18256,$\rho^{3}$ $Ari$       & 27,61      & 34,05           & -6,44               & 0,90           \\ \hline
  HD25998,$50 Per$               & 29,54      & 33,93           & -4,39               & 1,80           \\ \hline
  HD35296,$111 Tau$              & 29,44      & 33,78           & -4,34               & 3,00           \\ \hline
  HD39587,$\chi^{1} Ori$         & 29,08      & 33,69           & -4,61               & 3,60           \\ \hline
  HD75332                        & 29,56      & 33,87           & -4,31               & 2,90           \\ \hline
  HD82885,$SW Lmi$               & 29,3       & 33,42           & -4,12               & 3,30           \\ \hline
  HD115383,$59 Vir$              & 29,51      & 33,7            & -4,19               & 2,00           \\ \hline
  HD120136,$\tau$ $Boo$          & 28,95      & 33,87           & -4,92               & 1,20           \\ \hline
  HD176095                       & 29,0       & 33,93           & -4,93               & 0,80           \\ \hline
  HD182572,$31 Aql$              & 27,59      & 33,42           & -5,83               & 0,60           \\ \hline
  HD185144, $\sigma$ $Dra$       & 27,61      & 33,4            & -5,79               & 0,70           \\ \hline
  HD190007                       & 27,81      & 33,1            & -5,29               & 3,30           \\ \hline
  HD131156, $\zeta$  $Boo$       & 28,9       & 33,42           & -4,52               & 4,40           \\ \hline
  HD157856                       & 29,21      & 33,93           & -7,72               & 1,00           \\ \hline
  HD129333, $EK Dra$             & 30,01      & 33,64           & -3,63               & 13,80          \\ \hline
  $VY Ari $                      & 30,5       & 33,6            & -3,10               & 21,80          \\ \hline
  $V775 Her $                    & 30,0       & 33,38           & -3,38               & 30,00          \\ \hline
  $LQ Hia $                      & 29,6       & 33,35           & -3,75               & 12,40          \\ \hline
  $V838 Cen $                    & 29,9       & 33,38           & -3,48               & 21,60          \\ \hline
  $AG Dor $                      & 29,8       & 33,33           & -3,53               & 9,20           \\ \hline
  $MS Ser $                      & 30,1       & 33,35           & -3,25               & 10,00          \\ \hline
  $OU Gem $                      & 29,3       & 33,4            & -4,10               & 7,90           \\ \hline
  $V833 Tau $                    & 29,8       & 33,1            & -3,30               & 24,30          \\ \hline
  $EQ Vir $                      & 29,4       & 33,1            & -3,70               & 8,50           \\ \hline
  $BY Dra $                      & 29,6       & 32,95           & -3,35               & 16,50          \\ \hline
  $CC Eri $                      & 29,5       & 32,93           & -3,43               & 24,70          \\ \hline
  $DK Leo $                      & 29,1       & 32,9            & -3,80               & 11,00          \\ \hline
  $V1005 Ori $                   & 29,2       & 32,9            & -3,70               & 6,80           \\ \hline
  $BF CVn $                      & 29,0       & 32,82           & -3,82               & 9,00           \\ \hline
  $DT Vir $                      & 29,2       & 32,82           & -3,62               & 12,20          \\ \hline
  $AU Mic $                      & 29,5       & 32,8            & -3,30               & 12,70          \\ \hline
  $FK Aqr $                      & 29,4       & 32,8            & -3,40               & 11,90          \\ \hline
  $V1396 Cyg $                   & 28,9       & 32,78           & -3,88               & 20,80          \\ \hline
  $AD Leo $                      & 28,8       & 32,70           & -3,90               & 5,30           \\ \hline
  $GT Peg $                      & 29,2       & 32,7            & -3,50               & 7,4            \\ \hline
  $VZ Cmi $                      & 28,4       & 32,62           & -4,22               & 11,40          \\ \hline
  $EV Lac $                      & 28,8       & 32,62           & -3,82               & 16,80          \\ \hline

\end{tabular}
\end{center}

Acknowledgements. The authors thank the RFBR grant 09-02-01010 for
support of the work. The authors thank M.A. Livsits (IZMIRAN) and
R.E. Gershberg (CrAO) for discussions regarding this paper.

\vskip12pt {\bf REFERECES}
\vskip12pt

1. R.E. Gershberg, Solar-type activity of main sequence stars,
Odessa, Astroprint (2002).

2. S.L.Baliunas, R.A.Donahue, W.H.Soon, Astrophys.J. {\bf 438},
 269 (1995).

3. J.H.M.M.Schmitt, C.Liefke, Astron.Astrophys. {\bf 417}, 651
(2004).

4. E.A. Bruevich, M.M. Katsova, and D.D. Sokolov, Astron. zh. {\bf
78}, 827 (2001).

5. I.Yu. Alekseev, R.E. Gershberg, Astron. zh.
 {\bf 73}, 589 (1996).

6. I.Yu. Alekseev, Low-mass spotted stars, Odessa, Astroprint
 (2001).

7. J.T.Wright, G.W.Marcy, R.P.Butler, S.S.Vogt,
Astrophys.J.Suppl.Ser. {\bf 152}, 261 (2004).

8. R.R.Radick, C.W.Lockwood, B.A.Skiff and S.L.Baliunas,
Astrophys.J.Suppl.Ser. {\bf 118}, 239 (1998).

9. I.Yu. Alekseev, R.E. Gershberg, M.M. Katsova, M.A. Livshits,
Astron. zh. {\bf 78}, 558 (2001).

\bigskip

\end{document}